\documentclass[journal]{IEEEtran}

\usepackage{graphicx}
\usepackage{multicol}
\usepackage{booktabs}
\usepackage{graphicx}
\usepackage{subfigure}
\usepackage{float}
\usepackage{standalone}
\usepackage{amsmath}
\usepackage[linesnumbered,lined,ruled,commentsnumbered]{algorithm2e}
\usepackage{cite}
\usepackage{comment}
\usepackage{balance}
\usepackage{color}
\usepackage[utf8]{inputenc}
\usepackage{enumitem}
\usepackage{ragged2e}
\usepackage{verbatim}
\usepackage{amssymb}
\usepackage{hyperref}
\usepackage{multirow}
\usepackage{graphicx}
\usepackage[normalem]{ulem}
\usepackage[dvipsnames]{xcolor}
\usepackage{soul}

\soulregister\ref7
\soulregister\cite7
\soulregister\pageref7

\begin{document}

\title{Channel Estimation for One-Bit Multiuser Massive MIMO Using Conditional GAN}

\author{Yudi~Dong,~\IEEEmembership{Student Member,~IEEE,}
        Huaxia~Wang,~\IEEEmembership{Member,~IEEE,}
        and~Yu-Dong~Yao,~\IEEEmembership{Fellow,~IEEE}\vspace{-2em}
\thanks{Y. Dong and Y. D. Yao are with the Department of Electrical $\&$ Computer Engineering, Stevens Institute of Technology, Hoboken, New Jersey 07030, USA (e-mail: ydong6@stevens.edu;yyao@stevens.edu).

H. Wang is with the College of Engineering, Architecture and Technology (CEAT), Oklahoma State University, Stillwater, OK 74078 USA (e-mail: huaxia.wang@okstate.edu).}}

\maketitle

\begin{abstract}
Channel estimation is a challenging task, especially in a massive multiple-input multiple-output (MIMO) system with one-bit analog-to-digital converters (ADC). 
Traditional deep learning (DL) methods, that learn the mapping from inputs to real channels, have significant difficulties in estimating accurate channels because their loss functions are not well designed and investigated.
In this paper, a conditional generative adversarial networks (cGAN) is developed to predict more realistic channels by adversarially training two DL networks. 
cGANs not only learn the mapping from quantized observations to real channels but also learn an adaptive loss function to correctly train the networks.
Numerical results show that the proposed cGAN based approach outperforms existing DL methods and achieves high robustness in massive MIMO systems.
\end{abstract}


\section{Introduction}
\label{sec:intro}

Massive multiple-input multiple-output (MIMO) is a key technology to improve the system capacity and spectrum utilization in 5G wireless communications systems~\cite{larsson2014massive}. 
By deploying a large scale of antennas in base stations (BS), massive MIMO not only enhances the capability to reuse spectrum resources among multiple users but also significantly improves the data transmission rate due to the great anti-interference ability. 
However, current massive MIMO systems are typically equipped with high resolution analog-to-digital converters
(ADC), which results in high power consumption and hardware complexity. 
To address this issue, massive MIMO with one-bit ADCs is recommended to be an alternative solution~\cite{li2017channel}. 
The challenge is that such one-bit ADCs are accompanied by a crucial problem that the accurate channel estimation becomes more challenging due to the heavily quantized received signals from low-resolution ADCs.

Compressed sensing-based~\cite{vila2013expectation,mo2014channel,wang2019channel} approaches have been proposed to perform channel estimation in one-bit massive MIMO.
However, these channel estimators either rely on non-linear optimization algorithms with high complexity or have inadequate performance.
Recently, deep learning (DL) ~\cite{yang2019deep,he2018deep,huang2018deep,soltani2019deep, chun2019deep, gao2019deep,zhang2019deep} has been investigated in the channel estimation for massive MIMO  and achieves attractive success. 
For instance,  Yang~\textit{et al.}~\cite{yang2019deep} exploit multilayer perceptrons (MLPs) to learn the uplink-to-downlink mapping function in a massive MIMO system. 
However, MLPs are computationally expensive and hard to be trained in complicated data.
Hence, more papers~\cite{he2018deep,huang2018deep,soltani2019deep,  chun2019deep} using deep convolutional neural networks (CNNs)  have been presented to approximate a mapping between transmitted signals and channel statistics in massive MIMO. 
Similarly, some researches investigate massive MIMO with one-bit\cite{zhang2019deep} or low-resolution~\cite{gao2019deep} ADCs using deep learning, where deep MLPs or CNNs are utilized to estimate the channel from quantized received signals.

It is worth noticing that, using existing DL-based methods~\cite{yang2019deep, he2018deep, huang2018deep, soltani2019deep, chun2019deep, gao2019deep,zhang2019deep}, it is hard to generate a more realistic channel matrix due to the information loss with successive layers in neural networks~\cite{gabrie2018entropy}.
Neural networks could be suitable for classification or recognition problems where the output is a label and the information loss may not affect the performance.
But for the data generation problem (e.g., channel estimation), the information loss may lead to poor performance, which must be taken seriously.
Hence, the loss functions of neural networks have to be well designed and investigated to reduce information loss during the learning process when dealing with data generation problems.
However, current work of channel estimation either does not investigate loss functions~\cite{he2018deep,gao2019deep} or empirically uses a general loss function (i.e., $\mathcal{L}_{1}$ or $\mathcal{L}_{2}$ loss)~\cite{yang2019deep,huang2018deep,soltani2019deep,chun2019deep,zhang2019deep}.
The loss functions in these works are not well designed for the problem of channel estimation in massive MIMO systems, which could limit the performance to a large extent and lead to poor channel estimation results. 
Especially under the low signal-to-noise ratio (SNR) environment, an improper loss function could result in false optimization during the training phase.

In this paper, we propose to exploit conditional generative adversarial networks (cGANs) to solve the problem of channel estimation by introducing the adversarial model. 
In cGANs, the generative model (i.e., generator) and the adversarial model (i.e., discriminator) are trained adversarially to provide an adaptive loss according to different tasks and datasets, which can be called GAN loss~\cite{mirza2014conditional}. Compared with non-cGAN CNNs with fixed loss functions, the cGANs' architecture is more robust and easier to generate more realistic channels.
More specifically, we explore a cGAN based DL channel estimation framework for a massive MIMO system with one-bit ADCs. 
In a generator, we design an encoder-decoder CNN with a U-Net structure~\cite{ronneberger2015u} to estimate channel matrices from highly quantized observations.
The discriminator, which is a regular CNN, is utilized to evaluate the quality of estimated channels from the generator. 
By applying the adversarial training, the parameter update for the generator is not directly from the data but from the back-propagation of the discriminator, which attempts to learn the real distribution from groundtruth channels without the sophisticated design of loss functions.
The contributions of this paper are summarized as follows.
\begin{itemize}
  \item We propose a cGAN based DL approach in channel estimation for one-bit multiuser massive MIMO. This novelty exploits the cGAN architecture to address the intractable problem of loss function design in traditional DL methods, which makes estimated channels more accurate and realistic. 
  \item The GAN loss introduced by the cGAN structure plays a certain role in modifying the optimization of neural networks, so that the channel estimation system maintains good performance even in the environment of low SNRs.
  \item Numerical results show that the proposed cGAN based channel estimation approach outperforms existing DL-based methods and non-DL compressed sensing based methods, even using very short pilot sequence. The proposed approach is more robust in varying scenarios (e.g., different SNRs, different lengths of pilot sequences, and different numbers of BS antennas).
\end{itemize}


\section{System Model}
\label{sec:sys}

\begin{figure}[t]
\centering
\includegraphics[width=0.8\linewidth]{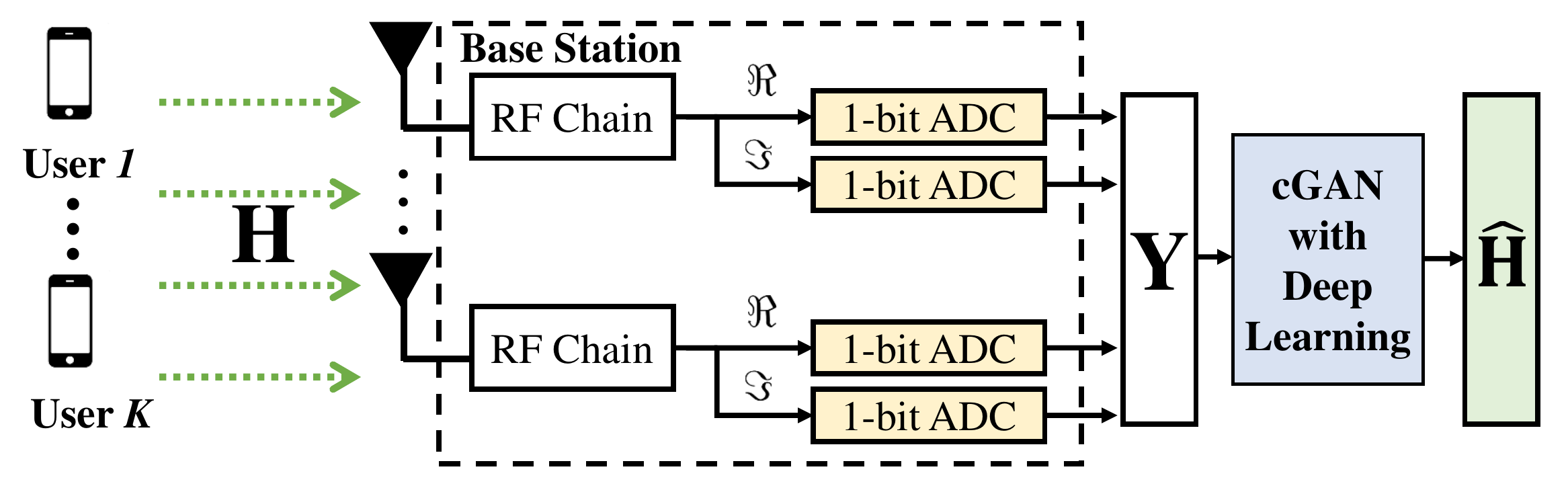}
\caption{Diagram of one-bit massive MIMO channel estimation.}
\label{fig:onebit}
\end{figure}

As shown in Fig.~\ref{fig:onebit}, we consider a signal-cell massive MIMO system with one-bit ADCs, where we have $K$ single-antenna equipped users and an $M$-antenna equipped BS. Each BS antenna is equipped with two one-bit ADCs.
The channels between the BS and users are generated using the accurate ray-tracing data obtained from Remcom Wireless InSite~\cite{alkhateeb2019deepmimo}, which is able to compute the comprehensive channel characteristics for each BS-user pair in each channel path. In particular, for $k^{\mathrm{th}}$ user in channel path $l$, we calculate the azimuth $\beta_{\textrm{azi}}^{k}$ and elevation angles of departure (AoDs) $\beta_{\textrm{aod}}^{k}$ at the BS, and the azimuth $\gamma_{\textrm{azi}
}^{k}$ and elevation angles of arrival (AoAs) $\gamma_{\textrm{aoa}}^{k}$ at the user side. 
Also, we compute the phase $\phi_{l}^{k}$, the receive power $P_{l}^{k}$, and the propagation delay $\lambda_{l}^{k}$ between the BS and $k^{\mathrm{th}}$ user in channel path $l$.

\begin{figure}[t]
	\centering
	\subfigure[]{
		\begin{minipage}{0.12\linewidth}
			\label{fig:pilot_ex}
			\centering
			\includegraphics[width=\linewidth]{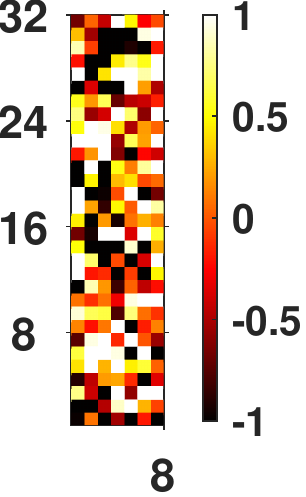}
		\end{minipage}
	}
	\subfigure[]{
		\begin{minipage}{0.27\linewidth}
			\label{fig:full_ex}
			\centering
			\includegraphics[width=\linewidth]{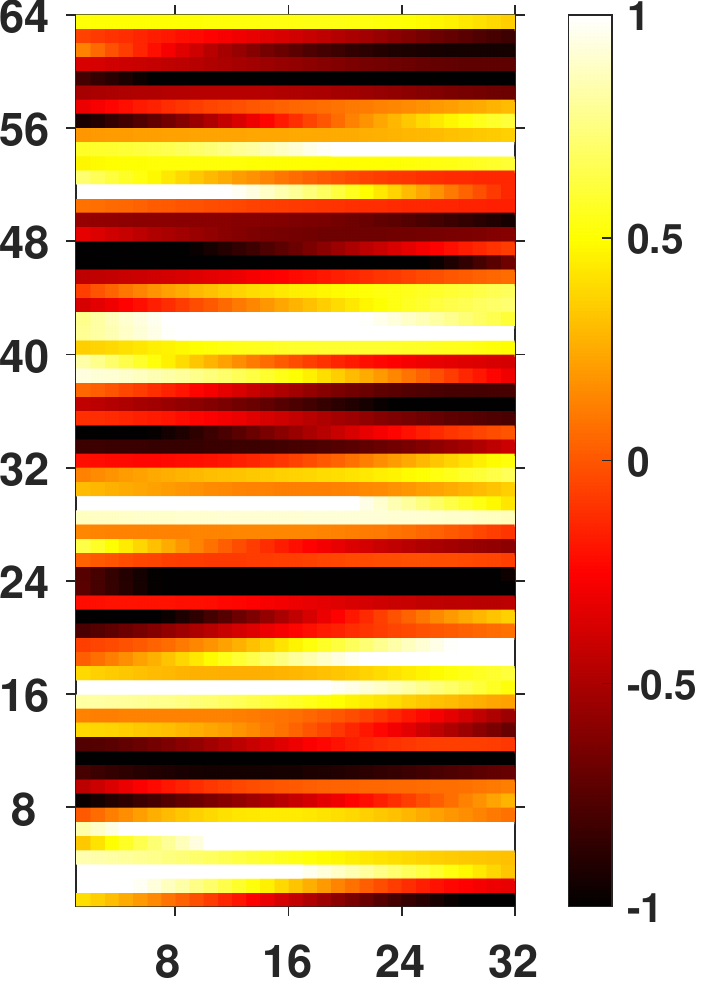}
		\end{minipage}
	}	
	\subfigure[]{
		\begin{minipage}{0.12\linewidth}
			\label{fig:qua_ex}
			\centering
			\includegraphics[width=\linewidth]{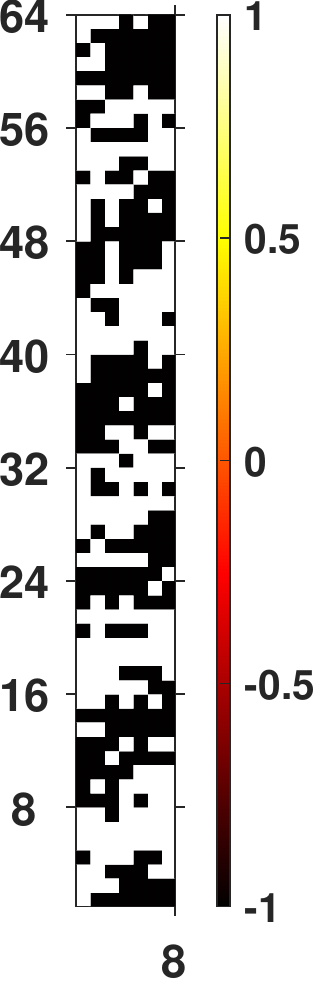}
		\end{minipage}
	}

\caption{Pseudo-color plots for real components of (a) pilot sequence $\Re (\mathbf{\Phi})$ (b) channel matrix $\Re (\mathbf{H})$ and (c) quantized measurements $\Re (\mathbf{Y})$.}
\label{fig:image_ex}
\end{figure}

Based on these channel parameters, we can construct channel vector $\mathbf{h}_{k}$ between the BS and user $k$,
\begin{equation}
\label{eq:channel_ve}
\mathbf{h}_{k} = \sum_{l= 1}^{L}\omega_{l}\cdot \mathbf{a}(\beta_{\textrm{azi}}^{k}, \beta_{\textrm{aod}}^{k})
\end{equation}
where $L$ is the number of channel paths. $\omega_{l}$ is the complex gain of each path $l$ and $\mathbf{a}(\beta_{\textrm{azi}}^{k}, \beta_{\textrm{aod}}^{k})$ denotes the array response of the BS, which are respectively expressed as
\begin{equation}
\omega_{l} = \sqrt{\frac{P_{l}^{k}}{K}}e^{j(\phi_{l}^{k} + \frac{2\pi k}{K}\lambda_{l}^{k} B)},
\end{equation}
\begin{equation}
\begin{split}
\mathbf{a}(\beta_{\textrm{azi}}^{k}, \beta_{\textrm{aod}}^{k}) = [&1, e^{jkd\cdot \mathrm{sin}(\beta_{\textrm{aod}}^{k})\mathrm{cos}(\beta_{\textrm{azi}}^{k})}, \cdots,\\
&e^{jkd(M-1)\cdot \mathrm{sin}(\beta_{\textrm{aod}}^{k})\mathrm{cos}(\beta_{\textrm{azi}}^{k})}]^{T},
\end{split}
\end{equation}
where $B$ is the system bandwidth and $d$ is the antenna spacing. 
Finally, the full channel matrix $\mathbf{H}$ for $K$ users is defined as
\begin{equation}
\label{eq:channel_matr}
    \mathbf{H} = [\mathbf{h}_{1}, \mathbf{h}_{2}, \cdots, \mathbf{h}_{k}, \cdots, \mathbf{h}_{K}]
\end{equation}
with dimension of $\mathbf{H}\in\mathbb{C}^{M\times K}$. 
The visualization of the channel matrix is depicted in Fig.~\ref{fig:full_ex} , where we plot the real component of a $64\times 32$ channel matrix with $64$ BS antennas and $32$ users. The figure colors correspond to the data values of the channel matrix and the colorbar on the right indicates the mapping of data values into the colors.

\subsection{Channel Estimation}
We perform channel estimation at the BS by using the pilot signals from users.
As depicted in Fig.~\ref{fig:onebit}, $K$ users simultaneously transmit pilot sequences with the length of $\tau$ to the BS. Then, the received signal $\mathbf{Y}$ at the BS after one-bit quantization is expressed as
\begin{equation}
\mathbf{Y} = \textrm{sgn}(\mathbf{H}\mathbf{\Phi} + \mathbf{N})
\end{equation}
with dimension of $\mathbf{Y}\in\mathbb{C}^{M\times \tau}$, where $\mathbf{\Phi}\in\mathbb{C}^{K\times \tau}$ is the pilot sequence from $K$ users and each user's pilots in $\mathbf{\Phi}$ is mutual orthogonal.
Fig.~\ref{fig:pilot_ex} shows the real components of a orthogonal pilot sequence from $32$ users with $8$ pilots for each user.
$\mathbf{N}\in\mathbb{C}^{M\times \tau}$ is a noise matrix at the BS which are drawn from a Gaussian distribution. The signum function $\textrm{sgn}(\cdot)$ is an element-wise operator for one-bit quantization defined as
\begin{equation}
\textrm{sgn}(x)= 
             \begin{cases}
             1 & ,x\geq 0  \\
             -1 & ,\textrm{otherwise} 
             \end{cases}
\end{equation}
which is applied to both the real and imaginary part of the argument $x$. 
Thus, $\mathbf{Y}$ is a quantized signal where its element takes values from the set $\left \{  1+j,1-j,-1+j,-1-j\right \}$. Fig.~\ref{fig:qua_ex} visually shows the real part of received signal $\mathbf{Y}\in\mathbb{C}^{64\times 8}$ after the one-bit quantization, where we have $M=64$ BS antennas and $K=32$ users using the pilot sequences with the length of $\tau = 8$. We can see that $\mathbf{Y}$ is a very low-resolution measurement.  

The objective of this paper is to recovery the channel matrix $\mathbf{\widehat{H}}$ from the highly quantized observations $\mathbf{Y}$ and the known pilot sequences $\mathbf{\Phi}$ using the cGAN trained model, which is a more accurate and robust approach to estimate a realistic channel matrix by using adversarial deep learning models.

\section{Channel Estimation for One-bit MIMO via cGAN}
\label{sec:gan}

We consider the received signal $\mathbf{Y}$, pilot sequence $\mathbf{\Phi}$, and channel matrix $\mathbf{H}$ as two-channel images with the dimension of $M\times \tau \times 2$, $K\times \tau \times 2$, and $M \times K \times 2$, respectively.  
Two channels of the image represent the real and imaginary part of a complex matrix.
Then, we can regard the channel estimation problem as an image-to-image translation problem, where we need to translate a low-resolution image with quantized $\mathbf{Y}$ to a high-resolution image with full channel matrix $\mathbf{H}$. 
In this paper, we adopt the cGAN
\footnote{Source Codes: \url{https://github.com/YudiDong/Channel_Estimation_cGAN}}
with deep neural networks to accomplish this channel estimation task.

\begin{figure}[t]
\centering
\includegraphics[width=\linewidth]{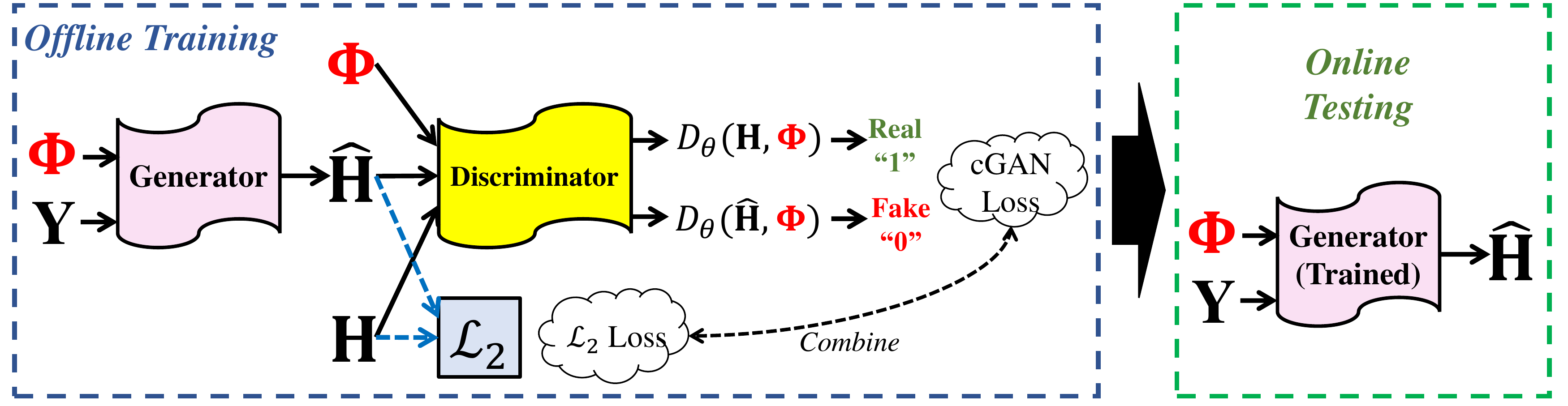}
\caption{Illustration of the proposed cGAN approach.}
\label{fig:cgan_f}
\end{figure}

The regular GAN is an architecture for training a generative model (i.e., generator) based on an adversarial model (i.e., discriminator). The generator learns a mapping from random noises to the real data. But this mapping has instability and randomicity. 
Therefore, cGANs~\cite{mirza2014conditional} are proposed as the extension of GANs, which learn a mapping from the conditional input to the real data. 
In our work, we utilize cGAN to learn the mapping relationship from received signals $\mathbf{Y}$ and pilots $\mathbf{\Phi}$ to real channel matrices $\mathbf{H}$. 
As shown in the Fig.~\ref{fig:cgan_f}, there are two neural networks respectively served as the generator and the discriminator in the offline training. 
The generator is responsible for estimating channel matrices $\mathbf{\widehat{H}}$ from the conditional input (i.e., quantized observations $\mathbf{Y}$ and the pilot sequence $\mathbf{\Phi}$). 
The discriminator can recognize a given input as either real label ``1" (i.e., drawn from the ground-truth $\mathbf{H}$ and the pilot sequence $\mathbf{\Phi}$) or fake label ``0" (i.e., drawn from the synthesized $\mathbf{\widehat{H}}$) and the pilot sequence $\mathbf{\Phi}$.
Once the trained generator is obtained, we can utilize it to perform channel estimation based on the new input of $\mathbf{Y}$ and $\mathbf{\Phi}$.

\subsection{Objective Function}
The objective of the cGAN is to make the generator synthesize the most realistic channel matrix to fool the discriminator. 
Meanwhile the discriminator needs to learn not to be easily tricked.
Both networks counteract their opponent to obtain the optimal results.
To achieve this optimization, we apply the GAN loss which is expressed as
\begin{equation}
\label{eq:ad_loss}
\begin{aligned}
\mathcal{L}_{\textrm{GAN}}(G_{\psi}, D_{\theta}, \mathbf{Y}, \mathbf{H}, \mathbf{\Phi})  = & \mathbb{E}[\log_{}D_{\theta}(\mathbf{H},\mathbf{\Phi})]\\
& + \mathbb{E}[\log_{}(1-D_{\theta}(G_{\psi}(\mathbf{Y, \mathbf{\Phi}}))],
\end{aligned}
\end{equation}
where $G_{\psi}$ denotes the generator parameterized by $\psi$ that synthesizes the channel matrix $\mathbf{\widehat{H}}$ (i.e., $G_{\psi}(\mathbf{Y}, \mathbf{\Phi})$) as more similar as the groudtruth in $\mathbf{H}$, and $D_{\theta}$ is the discriminator parameterized by $\theta$ aiming to differentiate the generated channel $\mathbf{\widehat{H}}$ from the real channel $\mathbf{H}$ (i.e., real label ``1" or fake label ``0").
Hence, we let $G_{\psi}$ minimize the GAN loss against the adversarial $D_{\theta}$ aiming to maximize it, 
\begin{equation}
\min_{\psi}\; \max_{\theta} \: \mathcal{L}_{\textrm{GAN}}(G_{\psi}, D_{\theta}, \mathbf{Y}, \mathbf{H}, \mathbf{\Phi}).
\end{equation}

As shown in Fig.~\ref{fig:cgan_f}, to ensure the right direction of generator optimization, we add a $\mathcal{L}_{2}$ loss~\cite{isola2017image} to the GAN loss, which is expressed as

\begin{equation}
\mathcal{L}_{2} = \mathbb{E}[\left \| \mathbf{H} - G_{\psi}(\mathbf{Y},\mathbf{\Phi}) \right \|^2].
\end{equation}

Finally, our objective function is
\begin{equation}
\min_{\psi}\; \max_{\theta} \: \mathcal{L}_{GAN}(G_{\psi}, D_{\theta}, \mathbf{Y}, \mathbf{H}, \mathbf{\Phi}) + \mathcal{L}_{2}.
\end{equation}

\begin{figure}[t]
\centering
\includegraphics[width=\linewidth]{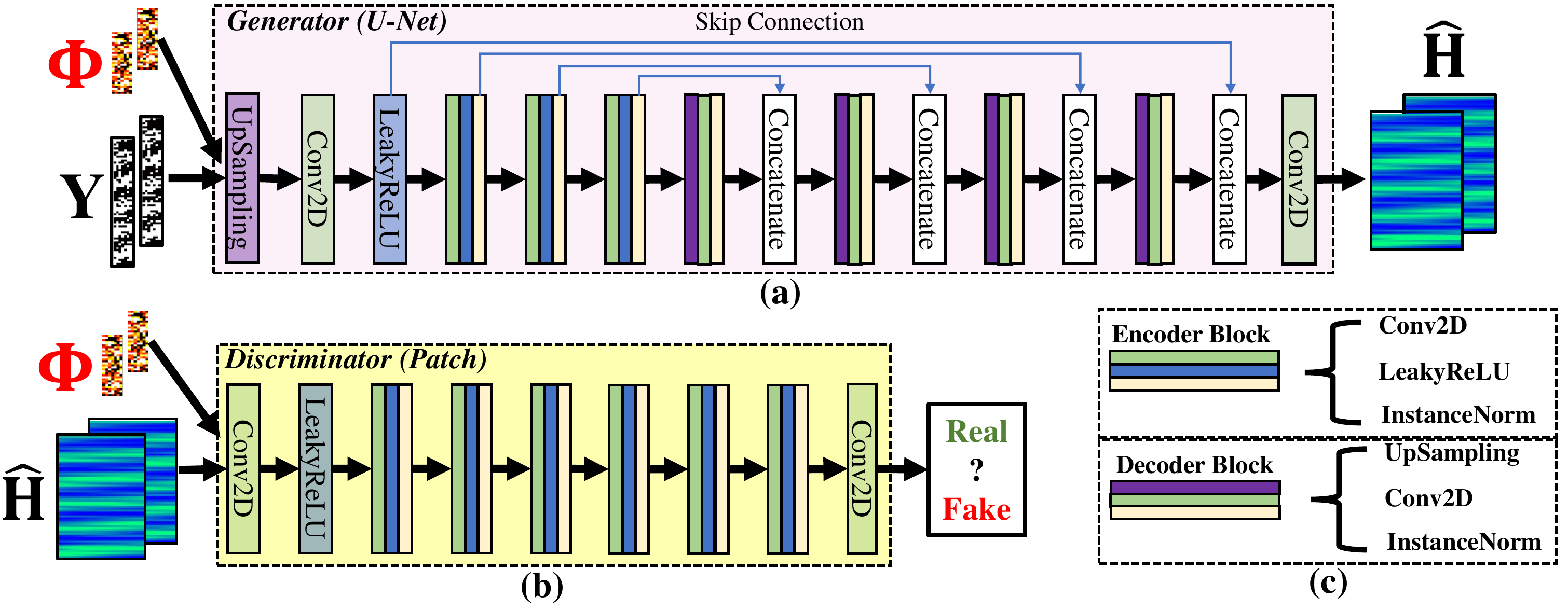}
\caption{Network architectures of the proposed cGAN approach. (a) Architecture of the generator. (b) Architecture of the discriminator. (c) Composition of a encoder block and decoder block.}
\label{fig:cgan}
\end{figure}

\subsection{Network Architecture}
We adopt U-Net~\cite{ronneberger2015u} architecture in the generator. 
U-Net is a full convolutional structure designed for image processing. 
As illustrated in Fig.~\ref{fig:cgan}, we first use one upsampling layer and one convolutional layer to rescale the input as the same size as $\mathbf{H}$. 
Then, we use three encoder blocks and four decoder blocks within U-Net architecture.
Each encoder block involves a convolutional layer, an instance normalization layer and a LeakyReLU activation layer. 
Each decoder block is composed of an upsampling layer, a convolutional layer and an instance normalization layer. We choose the instance normalization instead of batch normalization, which is able to accelerate the convergence of the generative model. 
For each convolutional layer, $128$ filters with the size of $4\times4$ are set as parameters. 
In addition, compared with the common encoder-decoder structure (i.e., auto-encoder), U-Net is distinguished by the addition of skip connection, where the feature maps of encoder blocks and decoder blocks are combined by concatenate layers to retain pixel level details at different resolutions.
Hence, the U-Net has an obvious effect on improving details, where the information of different scales could be preserved.
In addition, the generator may learn to produce samples with little variety to trick the discriminator, which causes mode collapse.
To avoid this problem, the output values of generator are normalized into $\left[-1,1\right]$ using hyperbolic tangent (tanh) activation function.

The discriminator is a straightforward convolutional neural network as shown in Fig.~\ref{fig:cgan}. 
In order to improve the recovery capability with detailed information, we use the patch architecture~\cite{isola2017image} in the discriminator to recognize the particulars of the input. 
Different from the regular discriminator that maps from the input to a single scalar output signifying real or fake, the patch discriminator maps the input to a receptive field, where each element signifies whether the part of the input is real or fake.  
To implement it, the front parts of the discriminator are composed of one convolutional layer, one LeakyReLU activation layer, and four encoder blocks. All convolutional layers are given as $512$ filters with the size of $4\times4$.
For the last layer, we substitute the fully connected layer for the convolutional layer to obtain the receptive field. Then, we average all responses of the receptive field to provide the ultimate output of the discriminator.

\section{Numerical Results}
\label{sec:eval}

In this section, we evaluate our cGAN based channel estimation approach in different scenarios by comparing with the U-Net without the cGAN architecture, the regular deep CNN approach used in MIMO systems~\cite{he2018deep,huang2018deep,soltani2019deep,chun2019deep}, and deep MLP approach used for one-bit channel estimation~\cite{zhang2019deep}.

\subsection{Dataset Preparation} 
The channel generation is based on the Wireless InSite ray-tracing that is used to build an indoor sub-6G massive MIMO with operating frequency of $2.5$ GHz~\cite{alkhateeb2019deepmimo}.
We generate four channel datasets with different numbers of BS antennas. 
Specifically, the numbers of BS antennas are set to $M= 64$, $M= 128$, $M= 192$, and $M= 256$. 
The number of users is fixed to $K = 32$.
All other parameters use the default settings, where the antenna spacing is the half wavelength, the bandwidth is $0.01$ GHz, and the number of multipaths is $L= 10$. 

Furthermore, we generate four channel datasets and each one contains $4200$ channel matrices $\mathbf{H}$ with the size of $64\times32$, $128\times32$, $192\times32$, and $256\times32$, respectively. In addition, the corresponding datasets of the received signal are generated based on the channel matrix datasets and pilot sequences using one-bit quantization. 
Meanwhile, we add noises with different SNRs to received signals. 

All datasets are divided into training, testing and validation sets by the ratios of $50$\%, $40$\%, and $10$\%, respectively.
To train the proposed cGAN model, the generator and the discriminator both apply MSProp algorithm~\cite{ruder2016overview} with a learning rate of $2\times 10^{-4}$ and $2\times 10^{-5}$, respectively. 

\subsection{Evaluation Metrics}
We utilize the normalized mean-squared-error (NMSE) to calculate the difference between the estimated matrix $\mathbf{\widehat{H}}$ and the real channel matrix $\mathbf{H}$, which is expressed as
$\textrm{NMSE} = 10\log_{10}\left \{  \mathbb{E}\left [   \frac{\left \|\mathbf{H} - \mathbf{\widehat{H}}\right \|^{2}}{\left \|\mathbf{H}\right \|^{2}} \right ] \right \}$,    
where $\left \| \cdot \right \|$ denotes the matrix norm calculation and $\mathbb{E}$ obtains values of expectation. And we calculate $10\log_{10}\left \{\cdot \right \}$ to obtain $\textrm{NMSE}$ values in decibels.

\subsection{Performance Analysis}

\begin{figure}[t]
	\centering
	\subfigure[]{
		\begin{minipage}{0.12\linewidth}
			\label{fig:ground}
			\centering
			\includegraphics[width=\linewidth]{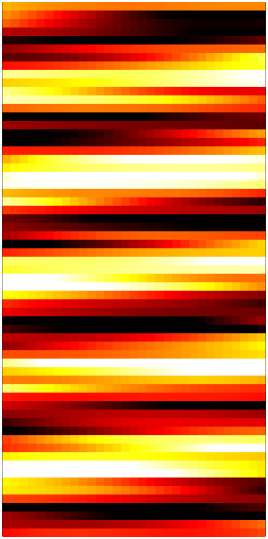}
		\end{minipage}
	}	
	\subfigure[]{
		\begin{minipage}{0.12\linewidth}
			\label{fig:cgan_re}
			\centering
			\includegraphics[width=\linewidth]{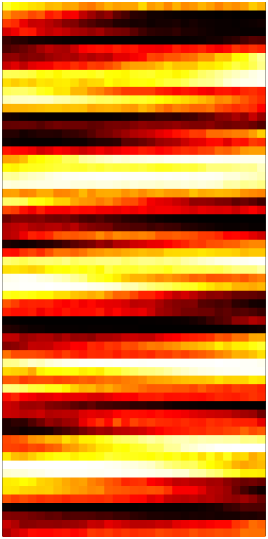}
		\end{minipage}
	}
	\subfigure[]{
		\begin{minipage}{0.12\linewidth}
			\label{fig:unet_re}
			\centering
			\includegraphics[width=\linewidth]{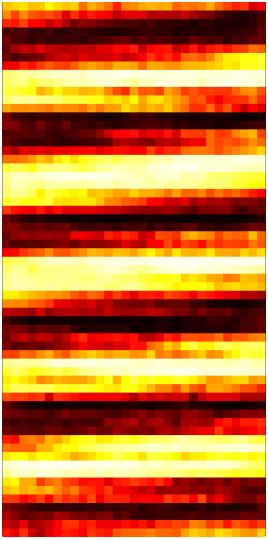}
		\end{minipage}
	}
		\subfigure[]{
		\begin{minipage}{0.12\linewidth}
			\label{fig:cnn_re}
			\centering
			\includegraphics[width=\linewidth]{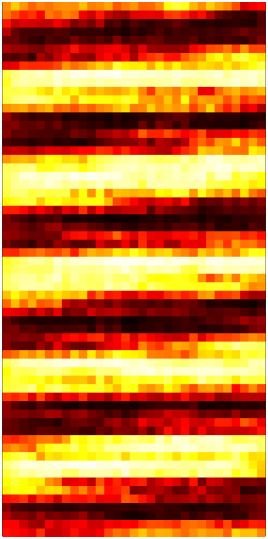}
		\end{minipage}
	}
	\subfigure[]{
		\begin{minipage}{0.12\linewidth}
			\label{fig:mlp_re}
			\centering
			\includegraphics[width=\linewidth]{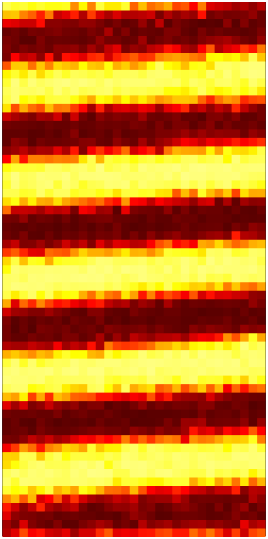}
		\end{minipage}
	}
\caption{Visualization of estimated channels with different methods. 
(a) Groundtruth $\Re (\mathbf{H})$.
(b) cGAN result $\Re (\mathbf{\widehat{H}})$.
(c) U-Net result $\Re (\mathbf{\widehat{H}})$.
(d) CNN result $\Re (\mathbf{\widehat{H}})$.
(e) MLP result $\Re (\mathbf{\widehat{H}})$.}
\label{fig:im}
\end{figure}

We compare our cGAN approach with other three deep learning methods, namely, U-Net, regular CNN~\cite{he2018deep, huang2018deep, soltani2019deep, chun2019deep}, and MLP~\cite{zhang2019deep}. 
All the DL models have the same number of layers and are trained efficiently with the same conditions.
Also, we compare our approach with the non-DL compressed sensing method (i.e., expectation-maximization Gaussian-mixture approximate message passing (EM-GM-GAMP)\cite{vila2013expectation}).
The result of U-Net presents the difference between the cGAN approach and the conventional training flow using $\mathcal{L}_{2}$ loss. 
The result of regular CNN and MLP reflects the performance difference without the cGAN structure and the U-Net architecture.

First, to intuitively see the performance improvements that the cGAN approach gains, we plot estimated channels and real channels as pseudo-color images in Fig.~\ref{fig:im}.
As we can see in Fig.~\ref{fig:ground} and Fig.~\ref{fig:cgan_re}, the images of the groudtruth channel and the cGAN generated channel are very similar, where the cGAN nicely produced the channel in details.  
For the U-Net, it leads to a somewhat blurry result as shown in Fig.~\ref{fig:unet_re}. 
Similarly, Fig.~\ref{fig:cnn_re} shows that the channel image is blurry using regular CNN. In Fig.~\ref{fig:mlp_re}, MLP also produces fuzzy channel results. 
We can see that with the cGAN architecture, it is capable to produce more realistic results.

\begin{figure}[t]
\centering
\includegraphics[width=0.75\linewidth]{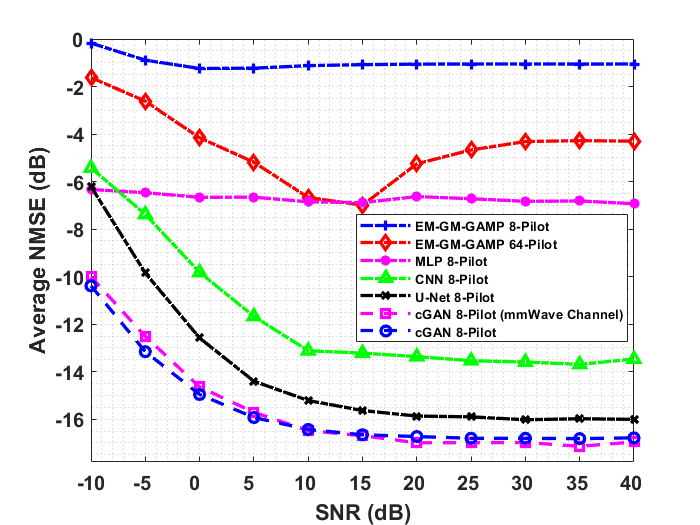}
\caption{Performance of different methods with varying SNRs.}
\label{fig:noise}
\end{figure}

Next, we investigate the effect of different SNRs on performance. 
Fig.~\ref{fig:noise} shows the NMSE performance of different methods with varying SNRs from $-10$ dB to $40$ dB.
We can see that all the DL methods outperform the compressed sensing method (i.e., EM-GM-GAMP).
Particularly, cGAN achieves the best performance across all SNR values, even under very short pilot length (i.e., 8 pilots) and under different channel realization (i.e., mmWave channel).
Remarkably, comparing with the U-Net, cGAN shows the great robustness at low SNR environments. 
For low SNR values varying from $-10$ dB to $0$ dB, the cGAN outperforms the U-Net with $3$ dB gains. That is due to the fact that the design of GAN loss and the cGAN architecture leads to a good optimization of the generator, which reduces the noise impact.
In addition, the performance of regular CNN is slightly worse than U-Net. MLP remains approximately consistent in NMSE performance at different SNRs, where NMSE values are around $-6$ dB.
For the EM-GM-GAMP method, it presents the worst results using $8$ pilots. Even by increasing the pilot length to $64$, the performance of EM-GM-GAMP still remains low and it also has the stochastic resonance phenomenon~\cite{rao2019channel, choi2019robust} where the estimation error increase at the SNRs greater than $15$ dB.

\begin{table}[t]
\centering
\caption{Comparison of Algorithm Complexity.}
\label{tab:time}
\resizebox{0.35\textwidth}{!}{%
\begin{tabular}{|c|c|c|l|l|c|}
\hline
\multirow{2}{*}{\begin{tabular}[c]{@{}c@{}}SNR\\ (dB)\end{tabular}} & \multicolumn{5}{c|}{\textit{Computation Time (ms)}} \\ \cline{2-6} 
             & \textit{cGAN}        & \textit{U-Net} & CNN   & MLP   & \textit{EM-GM-GAMP} \\ \hline
\textit{-10} & {\textbf{25.89}} & 26.86          & 20.43 & 22.34 & 823.41              \\ \hline
\textit{0}   & {\textbf{25.88}} & 27.83          & 20.28 & 23.06 & 599.26              \\ \hline
\textit{20}  & {\textbf{26.88}} & 26.82          & 20.77 & 22.45 & 531.38              \\ \hline
\end{tabular}%
}
\end{table}

\begin{figure}[t]
\centering
\includegraphics[width=0.75\linewidth]{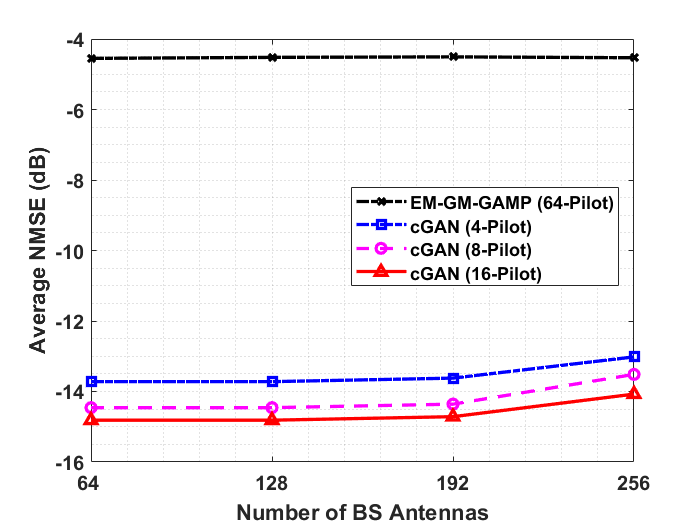}
\caption{Performance with different MIMO sizes and pilot lengths.}
\label{fig:size}
\end{figure}

In order to evaluate the algorithm complexity of cGAN, Table~\ref{tab:time} shows the computation time of deep learning models and the compressed sensing algorithm to perform channel estimation using a pilot signal. The cGAN and U-Net have the similar calculation time (i.e., $25$ ms) as the cGAN generator also adopts the U-Net architecture. Such computational complexity is acceptable in real-time transmissions for practical communications systems.
The CNN and MLP consume less time due to their simple architecture and less layers. The EM-GM-GAMP consumes more time than all deep learning methods and its computation time also increases at the low SNRs.
Finally, in Fig.~\ref{fig:size}, it shows that even with very small pilot size (i.e., $\tau < K$), the cGAN remains in good performance when more antennas are deployed at the BS. We can see that, with the pilot length of $4$ and $8$, the NMSE value of cGAN increases slightly when increasing the number of BS antennas from $64$ to $256$, but still stays good performance around $-14$ dB. Also, cGAN persists to achieve $-10$ dB gain over the EM-GM-GAMP method.
Moreover, we can see that the average NMSE difference between $4$ and $16$ pilots is not obvious. That is to say, the pilot length has no significant influence on the cGAN performance.


\section{Conclusion}
\label{sec:concl}
In this paper, we propose to apply conditional adversarial networks in the channel estimation for one-bit multiuser massive MIMO. 
These adversarial networks are able to adaptively learn a real loss from the data, which not only makes the model more robust but also makes the generated channels more realistic.
The evaluation results suggest that deep learning with conditional adversarial network is a more effective approach for channel estimation tasks, which significantly improves the channel estimation performance.

\bibliographystyle{IEEEtran}
\bibliography{ref}
\end{document}